\documentclass{aa}  
\usepackage{natbib}
\usepackage{lscape}
\usepackage{color}
\usepackage{soul}
\usepackage{graphicx}
\usepackage[varg]{txfonts}
\usepackage[pdftitle="The VEGAS survey"]{hyperref}

\begin{document} 

\title{On the assembly history of massive galaxies.}
\subtitle{A pilot project with VEGAS deep imaging and M3G integral field spectroscopy.}

   \author{Marilena Spavone 
          \inst{1}
          \and
          Davor  Krajnovi{\'c} \inst{2}
          \and
          Eric Emsellem \inst{3}
          \and
          Enrichetta Iodice \inst{1}
          \and
          Mark den Brok \inst{2}
          }
          
 \institute{INAF-Astronomical Observatory of Capodimonte, Salita Moiariello 16, I80131, Naples, Italy\\
              \email{marilena.spavone@inaf.it}
         \and
Leibniz-Institut für Astrophysik Potsdam (AIP), An der Sternwarte 16, 14482, Potsdam, Germany\\
\and
European Southern Observatory, Karl-Schwarzschild-Str. 2, 85748, Garching\\
             }

\date{Received ....; accepted ...}

\abstract{In this paper we  present the new deep images from the VEGAS survey of three massive 
($M_{*} \simeq 10^{12}$~M$_\odot$) galaxies from the MUSE Most Massive Galaxies (M3G) project, with distances in the range $151\leq D \leq 183$ Mpc:  
PGC007748, PGC015524 and PGC049940. 
The long integration time and the wide field of view of OmegaCam@VST allowed us to map the light 
and color distributions down to $\mu_g\simeq30$~mag/arcsec$^2$ and out to $\sim 2R_e$.
The deep data are crucial to estimate the contribution of the different galaxy's components, in particular the accreted fraction in the stellar halo.
The available integral-field observations with MUSE cover a limited portion of each galaxy (out to $\sim 1R_e$), but, from the imaging analysis we find that they map the 
kinematics and stellar population beyond the first transition radius, where the contribution of 
the accreted component starts to dominate.
The main goal of this work is to correlate the scales of the different components derived 
from the image analysis with the kinematics and stellar population profiles from the MUSE data. Results were used to address the assembly history of the three galaxies with the help of the theoretical predictions.
Our results suggest that PGC049940  has the lowest accreted mass fraction of 77\%.
The higher accreted mass fraction estimated for PGC007748 and PGC015524 (86\% and 89\%, respectively), combined with the flat 
$\lambda_R$ profiles suggest that a great majority of the mass has been acquired through major 
mergers, which have also shaped the shallower metallicity profiles observed at larger radii. 
}


\authorrunning{Spavone et al.}
\titlerunning{On the assembly history of the massive galaxies.}

\maketitle

\section{Introduction}

From both observational and theoretical side, it is broadly accepted that the stellar mass of
the present day massive (M$_{*} \simeq 3 \times 10^{11}$~M$_{\odot}$) and passive early-type galaxies (ETGs) results by the  
gradual accretion of satellites forming the extended stellar halo around a compact spheroid 
\citep[e.g.][]{Kormendy2009, vanDokkum2010, Huang2018, Spavone2017}. 
The accreted stellar content in ETGs, coming from the mass assembly in the stellar halo formed
the 'ex-situ' component, which mixed with the 'in-situ' component at small radii.
The in-situ stars have different properties in the light distribution, kinematics and 
stellar populations from those in the ex-situ region.

Simulations show that in the surface-brightness radial profiles of galaxies 
we may expect a {\it transition radius} ($R_{\rm tr}$), corresponding 
to variations in the ratio between the accreted ex-situ and the in-situ components \citep{Cooper2010,Deason2013,Amorisco2017}. The simulated profiles are, in fact, described by a first component identifying the stars formed in-situ, and by a second and a third component representing the relaxed and unrelaxed accreted stars, respectively. The radius marking the transition between each of these components is defined as transition radius.
Massive galaxies with a high accreted mass fraction are assumed to have a small first
$R_{\rm tr}$ \citep{Cooper2010,Cooper2013}. 
The outer stellar envelope should appear as a shallower exponential profile at larger radii, after a second transition radius, in the surface brightness distribution. 
Therefore, on the observational side, the study of the surface brightness profiles of 
ETGs out to the faintest levels turned out to be one of the main "tools" to quantify the 
contribution of the accreted mass. This method becomes particularly efficient when 
the outer stellar envelope starts to be dominant beyond the second transition radius
\citep{Seigar2007,Iodice2016,Iodice2017b, Spavone2017, Spavone2018, Spavone2020}.

To this aim, recent deep imaging surveys of nearby groups and clusters of galaxies have provided extensive analyses of the light and colour distribution of the brightest and massive members, out to the regions of the stellar halos where the imprints of the mass assembly reside \citep[e.g.][]{Duc2015, Capaccioli2015, Trujillo2016, Mihos2017, Spavone2018, Iodice2019a, Iodice2020}.

The VST Early-type GAlaxies Survey (VEGAS\footnote{http://www.na.astro.it/vegas/VEGAS/Welcome.html}, P.I. E. Iodice, \citealt{Capaccioli2015}) has recently occupied a pivotal role in exploring the low-surface brightness universe and 
studying galaxy properties as function of the environment. VEGAS data allowed to i) study the galaxy outskirts, detect the Intracluster Light (ICL) and Low Surface Brightness (LSB) features in the 
intra-cluster/group space \citep{Spavone2018, Cattapan2019, Iodice2020}; ii) trace the mass assembly in galaxies, by estimating the accreted 
mass fraction in the stellar halos and provide results that can be directly compared with predictions of galaxy formation models \citep{Spavone2017, Spavone2020}; iii) 
provide the spatial distribution of candidate globular clusters (GCs) in the galaxy outskirts and in the intra-cluster space \citep{Cantiello2018}; iv) detect 
ultra-diffuse galaxies (UDGs, \citealt{Forbes2019, Forbes2020, Iodice2020a}).

The stellar kinematics and population properties from the integrated light \citep[e.g.][]{Coccato2010, Coccato2011, Ma2014, Barbosa2018, Veale2018, Greene2019} 
and kinematics of discrete tracers like globular clusters (GCs) and planetary nebulae (PNe) \citep[e.g.][]{Coccato2013, Longobardi2013, Spiniello2018, Hartke2018, Pulsoni2020a, Pulsoni2020b} 
have also been used to
trace the mass assembly in the outer regions of galaxies.
In this field, key results are on the stellar population gradients out to the region of 
the stellar halos. They indicate a different star formation history in the central in-situ component from that in the galaxy outskirts
\citep[e.g.][]{Greene2015, McDermid2015, Barone2018, Ferreras2019}. Moreover, PNe and GCs have also been used as kinematic tracers and in some cases they show different kinematics than the central regions \citep{Coccato2009, Foster2013, Arnold2014}.

As the integral-field units become larger it is possible to map significant area of galaxies. A survey that mapped massive galaxies out to 2 effective radii is MUSE Most Massive Galaxies (M3G; PI: E. Emsellem) Survey. The main aim of this survey is to map the most massive galaxies in the densest galaxy environments at $z\sim0.045$ with the MUSE/VLT spectrograph \citep{Bacon2010}. M3G is a survey of 25 massive galaxies in dense environments, whose stellar velocity maps have been published in \citet{2018MNRAS.477.5327K}, and show that a large fraction of the galaxies in the sample have prolate-like rotation.

All the above mentioned findings, from both photometry and spectroscopy, are consistent 
with the theoretical prediction on the ETGs formation and evolution \citep{Cooper2013, Pillepich2018, Schulze2020, Cook2016}. 
In particular, the surface brightness and metallicity profiles appear shallower in the outskirts when repeated  mergers occur \citep{Cook2016}, i.e. the accreted fraction of metal-rich stars increases.

Moreover, by using {\it Magneticum Pathfinder} simulations, \citet{Schulze2020} find a correlation between the shape of the $\lambda_{R}$ and the galaxies' accretion history. In addition, they also found that the radius marking the kinematic transition between different galaxy's components provides a good estimate of the transition radius between the in-situ and accreted component in the photometric profiles.

The excellent quality of photometric and spectroscopic data from VEGAS and M3G, gives a unique opportunity to link the deep ($> 29$ mag/arcsec$^{2}$) optical photometry extending to several half-light radii ($R_e$) and the MUSE spectroscopic data (stellar kinematics and stellar populations) within the central $2 R_e$. 

The main goal of this work is to correlate the scales of different galaxies' components set from the deep photometry with the kinematics and stellar population properties derived from the integral field spectroscopy. 
While MUSE observations of massive galaxies in the nearby universe (within the M3G survey) provide spectroscopically derived constraints on their formation history (via stellar/gas kinematics and stellar populations), these are limited to the central 1-2 effective radii. Taking advantage of the extraordinary depth and extent of OmegaCam images from the VEGAS survey, we will investigate, for three Brightest Cluster Galaxies (BCGs) in M3G, the links between the central galaxy (spectroscopic) properties, and the large-scale (photometric) structures. The VEGAS surface brightness and colour profiles will serve a key structural parameterization of the main galaxy components, including the relaxed and non-relaxed stellar components, and an estimate of the accreted stellar mass. The MUSE data will then be used to reveal potential signatures of transitions between the structural components in stellar kinematics, angular momentum, age and metallicity distributions. The results, supported by detailed theoretical predictions, will help constraining the accretion histories of these targets.

This work is organized as follows. In Sec. \ref{data} we describe the data (both photometric and spectroscopic) used, and briefly describe the reduction steps. In Sec. \ref{analysis} we present the data analysis, while in Sec. \ref{sec:fit} we show the adopted procedure for fitting the surface brightness profiles. In Sec. \ref{sec:SP} we compare the observables derived from the surface photometry with the stellar population properties derived from MUSE data, and in Sec. \ref{sec:discussion} we compare observational results with the theoretical predictions and draw our conclusions.

\section{Observations and data reduction}\label{data}

In the following sections we describe the VST and MUSE observations and data reduction.

\subsection{Deep Photometry}
The photometric data presented in this work are part of the VEGAS survey \citep{Capaccioli2015}.
VEGAS is a multi-band {\it u}, {\it g}, {\it r} and {\it i} imaging survey carried out with the European Southern Observatory (ESO) Very Large Telescope Survey Telescope (VST). 
The VST is a 2.6~m wide field optical telescope \citep{Schipani2012} equipped with OmegaCAM, a $1^{\circ} \times 1^{\circ}$ camera with a resolution of $0.21$~arcsec~pixel$^{-1}$. 
The data we present in this work were acquired in the {\it g} and {\it r} bands, in visitor mode (run IDs: 100.B-0168(A), 101.A-0166(A), 102.A-0669(A)), 
during dark time in photometric conditions, with an average seeing of FWHM$\sim 1$~arcsec in the $g$ band and FWHM$\sim 0.7$~arcsec in the $r$ band. 
The total exposure time is 2.5 hours for each target and in each band. 

All the imaging data were processed by using the dedicated {\it AstroWISE} pipeline developed to reduce OmegaCam observations \citep{McFarland2013, Venhola2018}. 
The various step of the AstroWise data reduction were extensively described in \citet{Venhola2017,Venhola2018}. 

Given that the galaxies analyzed in this work have a angular extent much smaller than the VST field, observations for these objects have been performed by using the standard diagonal dithering strategy. As described in \citet{Capaccioli2015} and \citet{Spavone2017}, for the targets observed with this strategy, the background subtraction is performed by fitting a surface, typically a 2D polynomial, to the pixel values of the mosaic that are unaffected by celestial sources or defects. 

\subsection{2D Spectroscopy}

The spectroscopic data are taken from the M3G Survey that mapped 25 galaxies found in dense environments. The sample and the data reduction were presented in \citet{2018MNRAS.477.5327K}, and here we briefly outline the most relevant points. The M3G sample is built of galaxies brighter than -25.7 mag in 2MASS $K_s$-band, divided in two sub-samples comprising 11 galaxies in the three richest \citep{1989ApJS...70....1A} clusters of the Shapley Super Cluster and 14 brightest cluster galaxies (BCGs) selected from cluster that are richer than the Virgo Cluster (three of these are also in the Shapley Super Cluster sub-sample). This paper analyses the three BCGs found both in VEGAS and M3G samples. 

The data reduction was performed using the MUSE data reduction pipeline \citep{2020A&A...641A..28W}, following the standard steps: producing the master bias and flat-field calibration files, the trace tables, wavelength calibration files and line-spread function (LSF) for each slice. When available we also used twilight flats. Instrument geometry and astrometry files were provided by the GTO team for each observing run. We used separate sky fields to construct the sky spectra, associated with the closest in time on-target exposure. The response function was obtained from the observation of a standard star (for each night), but there were no observations of telluric standards. In addition to the reduction steps explained \citet{2018MNRAS.477.5327K}, for this paper we also removed the telluric lines using the MOLECFIT software \citep{2015A&A...576A..77S}, which computes a theoretical absorption model based on a radiative transfer code and an atmospheric molecular line database. The main contribution to the telluric absorptions comes from the oxygen either in the molecular form ($\gamma-$ and B-bands), or as ozone (Choppuis land), and water \cite[see Fig.~1 in][]{2015A&A...576A..77S}. MOLECFIT was applied to the full wavelength region of the MUSE data cubes. The final data cubes were obtained by merging all individual exposures.

\section{Data analysis}\label{analysis}

\subsection{Imaging}
\label{ss:iso}

We used sky-subtracted and stacked images of each galaxy in the {\it g} and {\it r} bands to perform the photometric analysis. 
We first estimate any residual fluctuations in the sky-subtracted images, therefore we obtain an estimate on the accuracy of our sky-subtraction. The procedure is the same adopted in previous VEGAS papers \citep{Capaccioli2015,Iodice2016,Spavone2017,Spavone2018}. In short, we extracted from the sky-subtracted images of each galaxy, after masking all the bright sources (galaxies and stars) and background objects, the azimuthally averaged intensity profile (by using the IRAF task ELLIPSE, \citealt{Jedrzejewski1987})  out to the edges of the frame by fixing both the position angle and the ellipticity of the galaxy. From this profile we estimated a residual background of $\sim 0.04 \pm 0.02$ counts in {\it g} and $\sim 0.10 \pm 0.06$ in {\it r}, by extrapolating the outer trend. From this procedure we also estimated the outermost radius where counts are consistent with the average background level. Such a limiting radius sets the surface brightness limit of the VST light profiles. The limiting radii ($R_{lim}$) are $\sim$ 3 arcmin ($\sim$ 160 kpc, $R_{lim}/R_{e} \sim 7$) for PGC007748, $\sim$ 10 arcmin ($\sim$ 438 kpc, $R_{lim}/R_{e} \sim 12$) for PGC015524 and $\sim$ 12 arcmin ($\sim$ 598 kpc, $R_{lim}/R_{e} \sim 19$) for PGC049948, in both {\it g} and {\it r} bands.

To account for the broadening effect of the seeing on the light distribution of galaxies, we used the Point Spread Function (PSF) derived in \citet{Capaccioli2015} by using VST images, and deconvolved each galaxy in our sample for the VST PSF, by using the Richardson-Lucy (hereafter RL) algorithm \citep{Lucy1974,Richardson1972}. The robustness and the reliability of this deconvolution technique are shown in \citet{Spavone2020}.

Before performing the analysis of the light distribution,
where needed, two-dimensional models of the bright stars close in projection to the galaxy under study, were derived and subtracted from the image.

\begin{figure*}
\includegraphics[width=10cm]{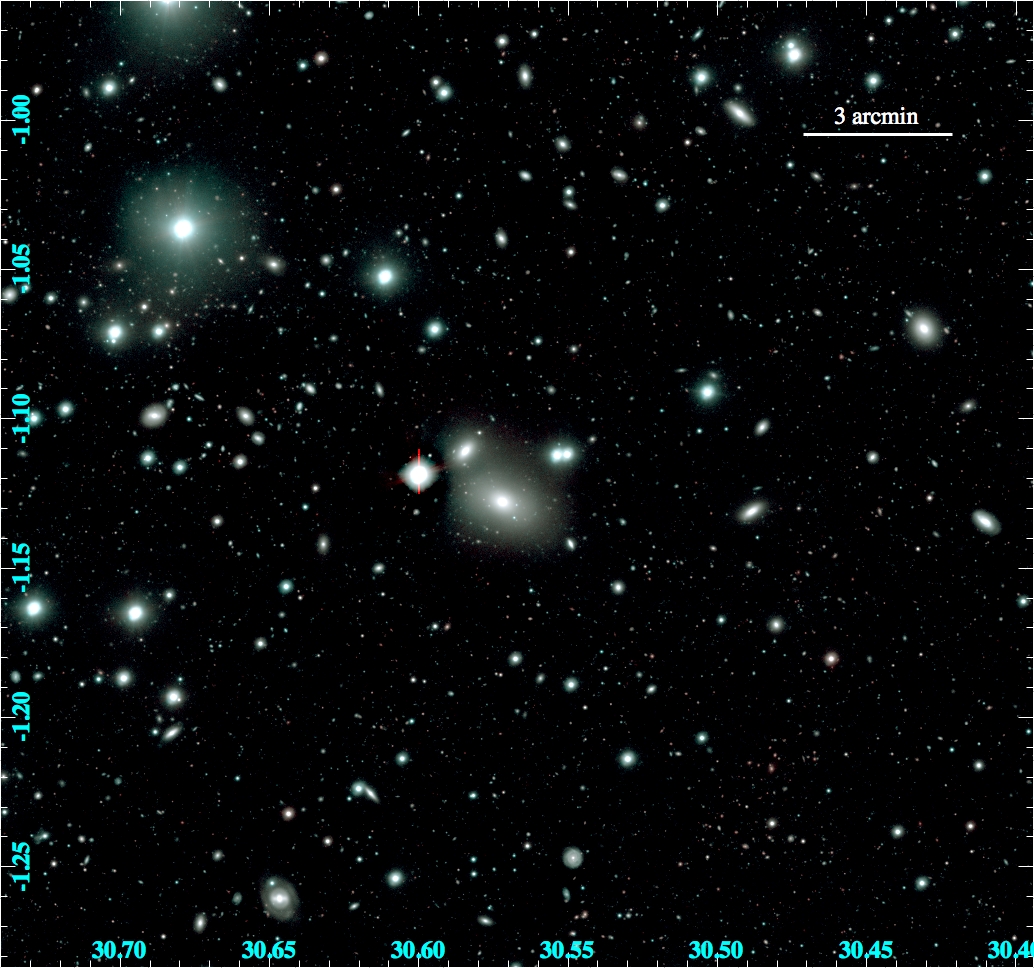}
\includegraphics[width=9cm]{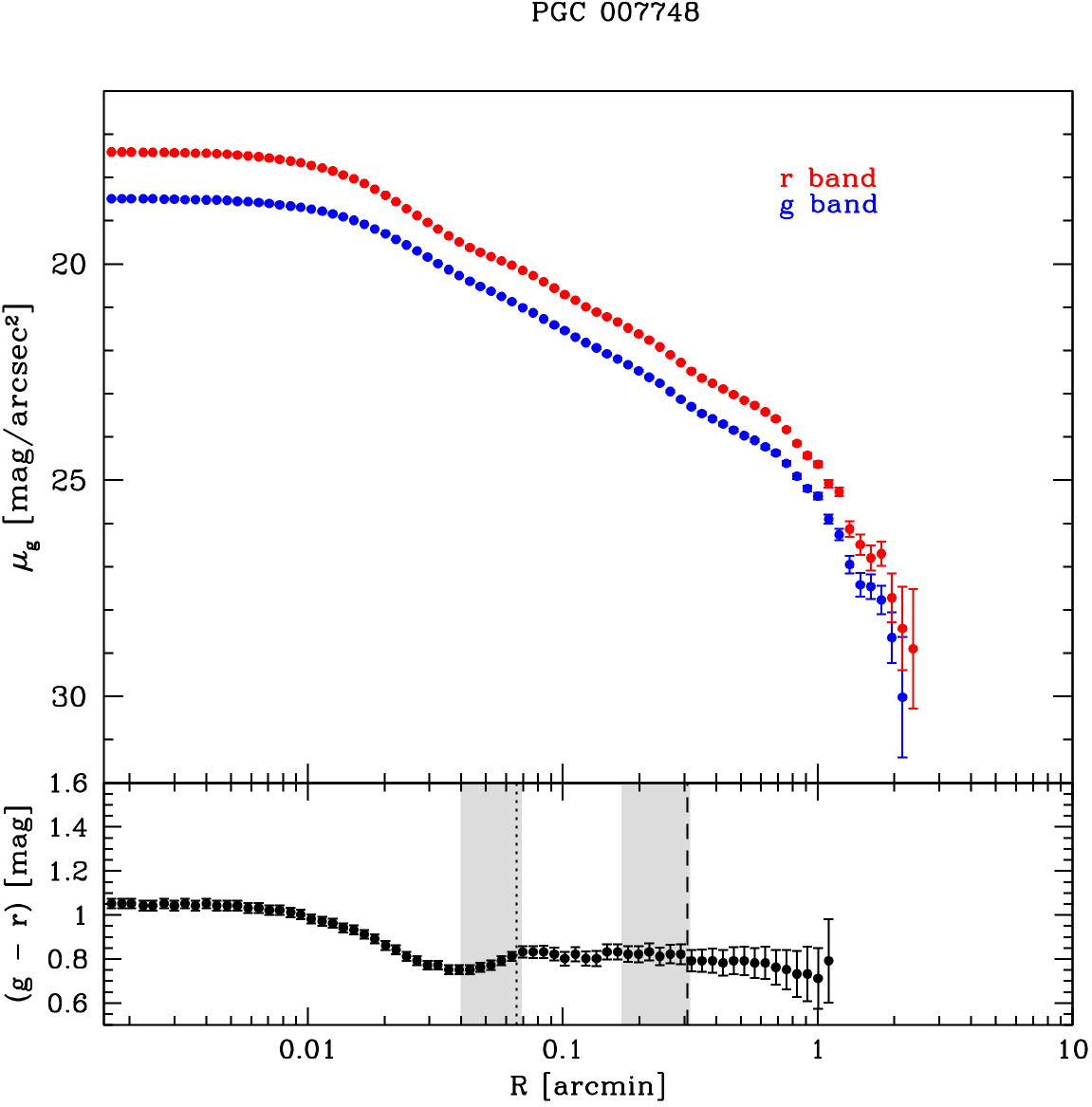}
\caption{{\it Left panel}: VST color-composite $gr$ image of the $20.7$ arcmin $\times\ 19.4$ arcmin field around PGC 7748. 
{\it Top-right panel}: Azimuthally averaged surface brightness profiles in the {\it g} (blue) and {\it r} (red) bands. {\it Bottom-right panel}: Azimuthally averaged ({\it g - r}) color profile. The vertical dotted and dashed lines mark the position of the first and second transition radii, respectively, while the grey shaded areas are the transition regions (see Sec. \ref{fit}).}
\label{PGC7748}
\end{figure*}

\begin{figure*}
\includegraphics[width=10cm]{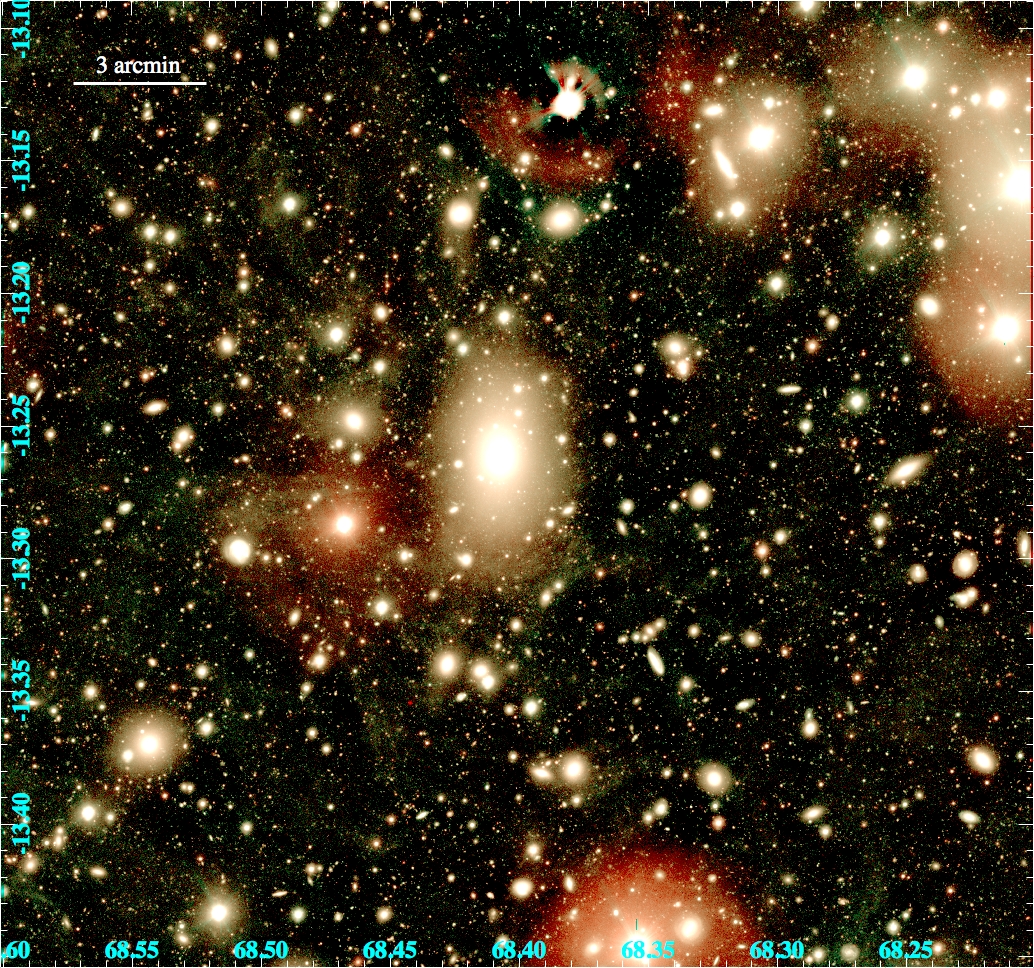}
\includegraphics[width=9cm]{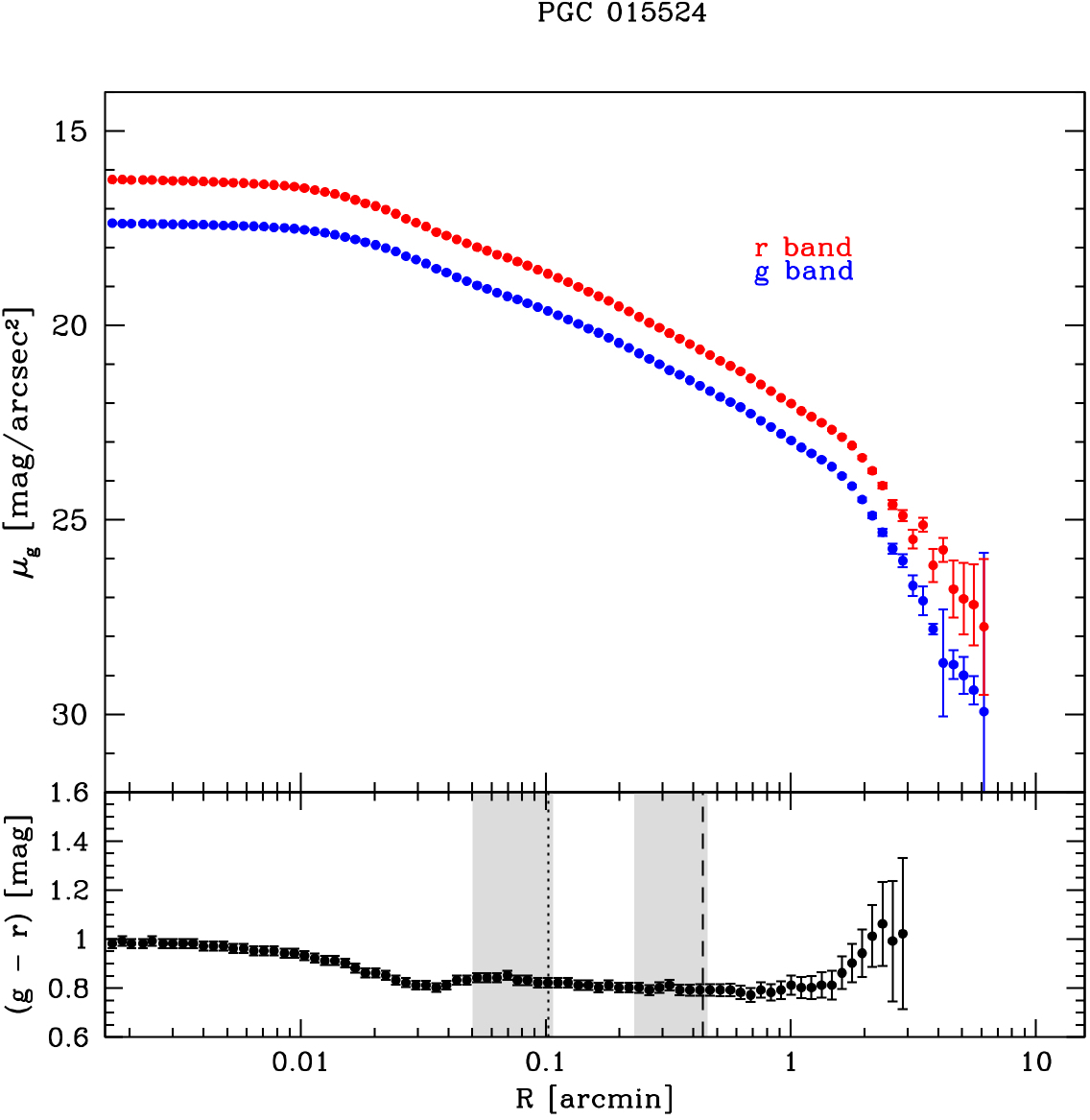}
\caption{Same as Fig.~\ref{PGC7748} for PGC 015524. The image size is $23.3 \times 21.9$~arcmin.}
\label{PGC015524}
\end{figure*}

\begin{figure*}
\includegraphics[width=10cm]{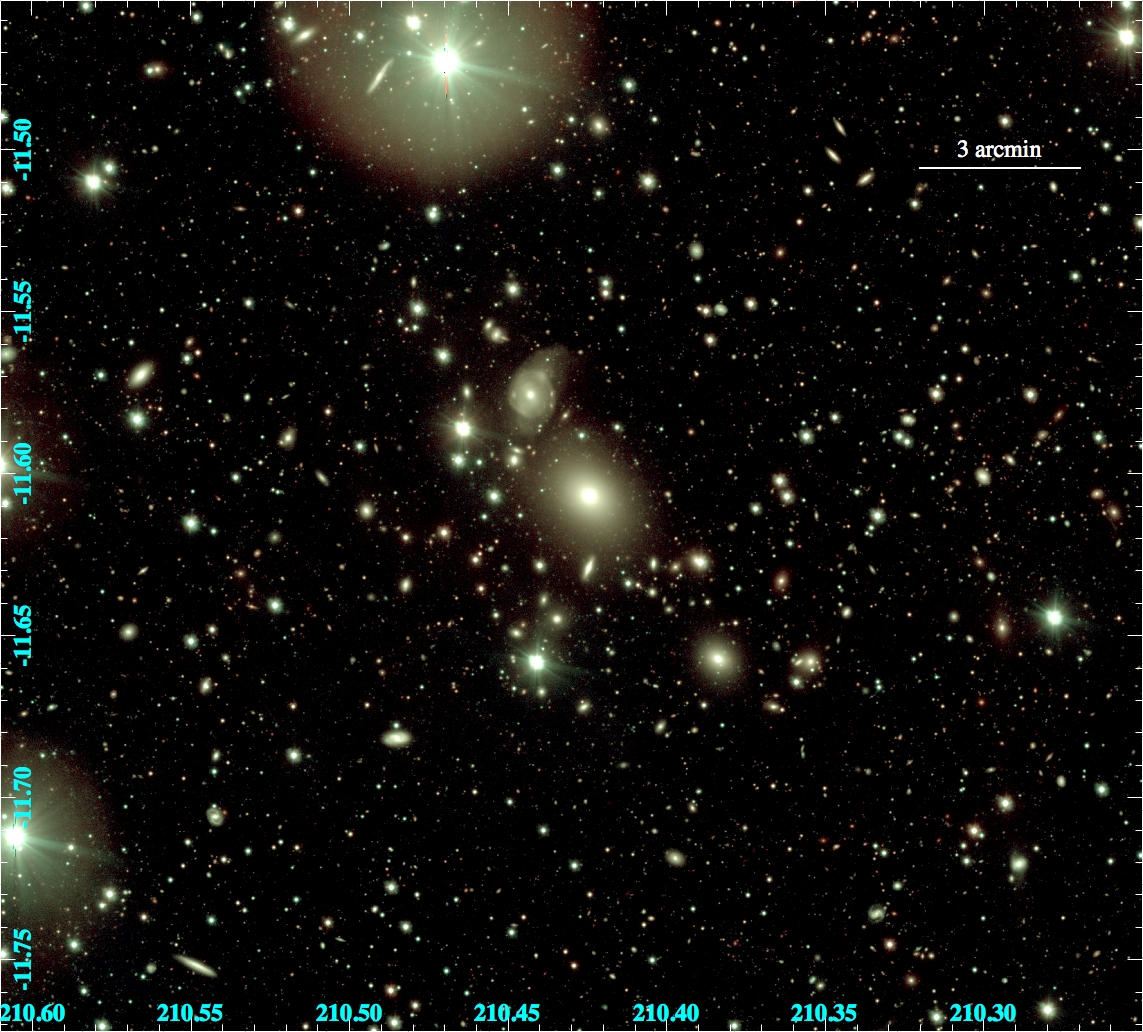}
\includegraphics[width=9cm]{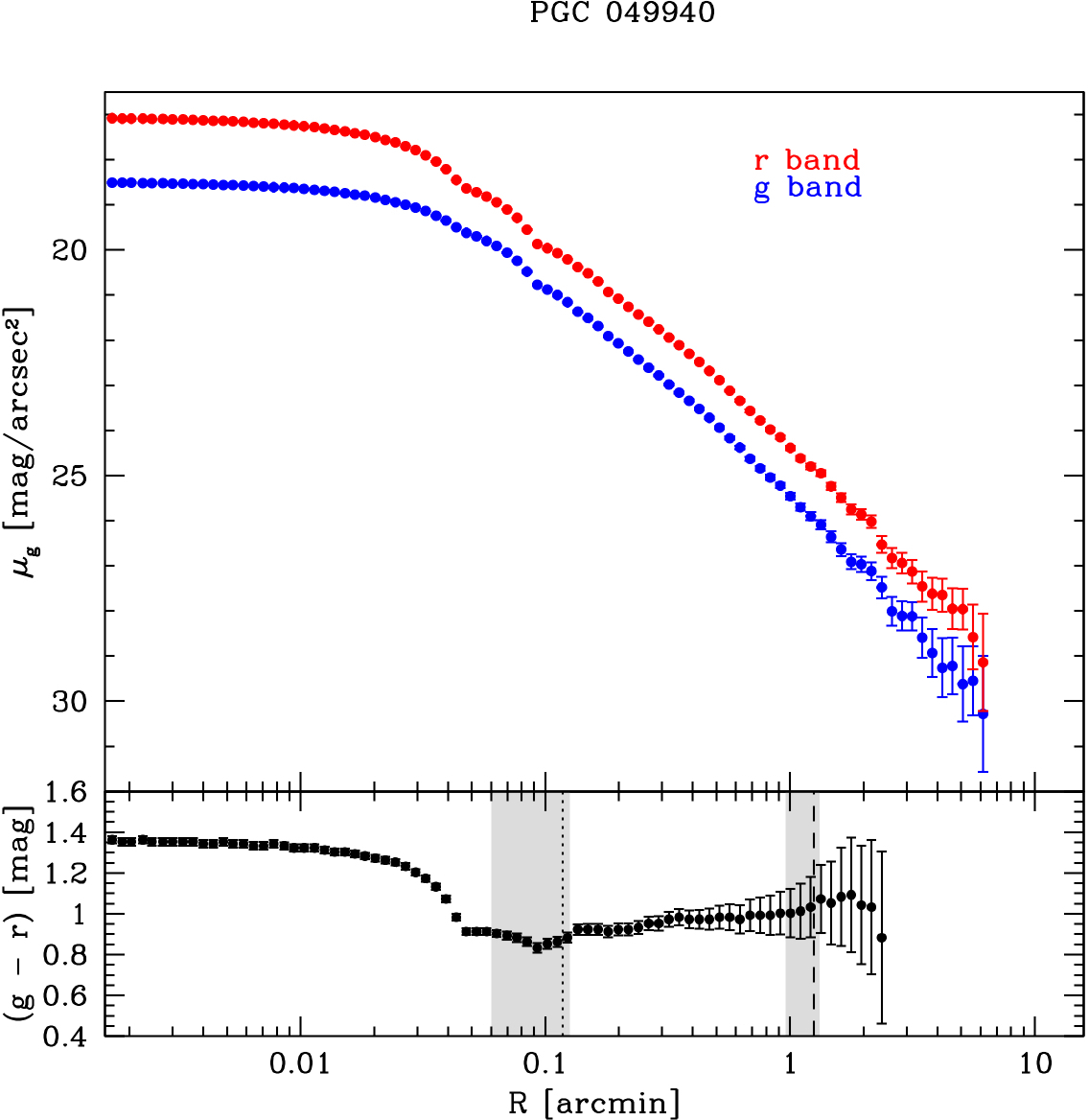}
\caption{Same as Fig.~\ref{PGC7748} for PGC 049940. The image size is $21.11 \times 19.13$~arcmin.}
\label{PGC049940}
\end{figure*}

The isophotal analysis has been performed with the IRAF\footnote{ IRAF (Image Reduction and Analysis Facility) is distributed by the National Optical Astronomy Observatories, which is operated by the Associated Universities for Research in Astronomy, Inc. under cooperative agreement with the National Science Foundation.} task ELLIPSE, which provides the geometrical parameters and the light distribution azimuthally averaged within isophotal annuli of specified thickness. The isophote fit was performed for each galaxy by masking all the bright sources in the field (stars and background galaxies). In this second ELLIPSE run both the ellipticity and position angle are left free. From this analysis we derived the ellipticity ($\epsilon$) and position angle (PA) radial profiles, the azimuthally averaged surface brightness profiles in each band and the azimuthally averaged {\it (g - r)} color profiles. The results are shown in Fig. \ref{PGC7748}, \ref{PGC015524} and \ref{PGC049940}, for PGC007748, PGC015524 and PGC049940 respectively.

The growth curves obtained by isophote fitting have been integrated and extrapolated to derive total magnitudes $m_{T}$, effective radii $R_{e}$, and corresponding effective magnitudes $\mu_{e}$ in each band (see Tab.\ref{par}).  

\begin{table*}
\setlength{\tabcolsep}{2.1pt}
\tiny
\begin{center}
\caption{Distances and photometric parameters for the
  sample galaxies.} \label{par}
\vspace{10pt}
\begin{tabular}{lcccccccccccccccc}
\hline\hline
Object & $D$ &$A_{g}$&$A_{r}$&$\mu_{e,g}$ &
$r_{e,g}$&$m_{T,g}$& $\mu_{e,r}$ &
$r_{e,r}$&$m_{T,r}$& log $M_{\ast}$\\
    &[Mpc] &[mag] &[mag] & [mag/arcsec$^{2}$] & [arcsec] &[mag] &[mag/arcsec$^{2}$] & [arcsec] &[mag]& [$M_{\odot}$]\\
&(a)&(b)&(b)&(c)&&(c)&(c)& &(c)& (d)\\
\hline \vspace{-7pt}\\
PGC 007748 & 183 $\pm$ 12  &   0.090$\pm$0.010 & 0.062$\pm$0.010&23.75$\pm$0.03 & 28.11 $\pm$ 0.10&13.62 $\pm$ 0.07 & 22.86 $\pm$ 0.02 & 26.14 $\pm$ 0.08 & 12.89 $\pm$ 0.06& 12.072 $\pm$ 0.140\\
PGC 015524 & 151 $\pm$ 30  &   0.447$\pm$0.010& 0.309$\pm$0.010&23.39 $\pm$ 0.03 & 45.27 $\pm$ 0.08&10.97 $\pm$ 0.04  & 20.39 $\pm$ 0.01& 49.78 $\pm$ 0.05&10.11 $\pm$ 0.04 &12.233$\pm$ 0.213 \\
 PGC 049940 & 171 $\pm$ 12  &   0.217$\pm$0.010& 0.150$\pm$0.010& 23.95 $\pm$ 0.04& 32.69 $\pm$ 0.30&13.11 $\pm$ 0.20 &23.27 $\pm$ 0.03 &37.41 $\pm$ 0.2 &12.11 $\pm$ 0.13 &12.078$\pm$ 0.166 \\
\hline
\end{tabular}
\tablefoot{\tablefoottext{a}{Distance of PGC 7748 is
  from \citet{Saulder2016}; distance of PGC 015524 is from
  \citet{Tully2013}; distance of PGC 049940 is from \citet{Fixsen1996}.}
\tablefoottext{b}{Sources for the extinction correction in the {\it g}
  and {\it r} band are \citet{Schlafly2011}.}
\tablefoottext{c}{Derived by integrating the growth curves and corrected for interstellar extinction.}
\tablefoottext{d}{Stellar masses are based on the virial estimates from \citet{2018MNRAS.477.5327K}, and their estimated uncertainties are driven by the distance and the size estimate.}}
\end{center}
\end{table*}

\subsection{Spectroscopy}

For the comparison with the imaging we are primarily interested in two sets of spectroscopic data products: stellar kinematics and properties of stellar populations. In particular, we want to characterise the radial profiles of the stellar velocity dispersion, stellar angular momentum, stellar ages and metallicities. 

Stellar kinematics was extracted in two steps. Based on the signal to noise ratio (S/N) of individual spectra, MUSE data cubes were binned using the Voronoi method \citep{2003MNRAS.342..345C}\footnote{\label{fn:cap} Available at http://purl.org/cappellari/software.} to a target S/N of 50. This was followed by extraction of kinematics based assuming a Gaussian line-of-sight velocity distribution, using the {\tt pPXF} software \citep{2004PASP..116..138C, 2017MNRAS.466..798C}\footnote{See footnote~\ref{fn:cap} for availability}. We masked the regions of possible emission-lines and residual sky lines, and fitted the wavelength region bluewards of 7000 \AA, applying a 12th order additive polynomial, which is appropriate for the long wavelength region used, as suggested by the analysis in Appendix A4 of \citet{2017ApJ...835..104V}. The wavelength limit is based on tests which showed that the extraction of kinematics in the blue part of MUSE spectra is fully consistent with kinematics of the SAURON survey \citep{2004MNRAS.352..721E}, while including the full MUSE spectrum can lead to inconsistencies \citep{2015MNRAS.452....2K}.  We masked all emission lines in that spectral range (i.e. H$\beta$ [OIII]$\lambda4959,5007$, [NI]$\lambda5197,5200$, [OI]$\lambda6300,6364$, [NII]$\lambda 6548,6583$, H$\alpha$ and [SII]$\lambda 6716,6731$) as well as any residual skylines (e.g. at 5557 \AA\, and around 6300 \AA). An example of a pPXF fit to M3G spectra, together with the fitting region and masked wavelenghts, is provided in \citet{2018MNRAS.477.5327K}. As templates we used the MILES library of stellar spectra \citep{2006MNRAS.371..703S, 2011A&A...532A..95F}, which were convolved to match the varying MUSE spectral resolution \citep{2017A&A...608A...5G} in the fitted spectral regions. 

Extraction of the stellar population parameters proceed similarly using the {\tt pPXF}, but this time using the E-MILES single stellar population (SSP) models \citep{2016MNRAS.463.3409V}. We used SSP models, based on the Padova isochrones \citep{2000A&AS..141..371G} and \citet{2001MNRAS.322..231K} initial mass function, with parameters distributed in a grid of log(age) between 0.1 and 14.1 Gyrs and metallicities [Z/H] between -1.71 and 0.22. The {\tt pPXF} fit assigns a weight to each of the SSP spectra, and the final age and metallicities were calculated for each bin as the mass-(or luminosity-)weighted averages of the SSP parameters (age and metallicity) using the weights assigned by the {\tt pPXF} fit. For the extraction of stellar populations parameters we applied only multiplicative polynomials of the 12th order, masked the emission-lines as outlined above, and used the same fitting region (4850-7000 \AA). Errors on the kinematic and line-strength parameters were obtained via 500 Monte-Carlo simulations, where the original spectrum was perturbed by a random value drawn from the standard deviation of the residuals of the pPXF fits. We derived both mass and luminosity-weighted metallicities and ages, but we present here only the luminosity-weighted parameters for the consistency with other (luminosity-weighted) parameters that we use (SB, $\lambda_{R}$ and $\sigma$). Nevertheless, we verified that the results do not change if we would use mass-weighted metallicities and ages.

\begin{figure*}
\center
\includegraphics[width=18cm]{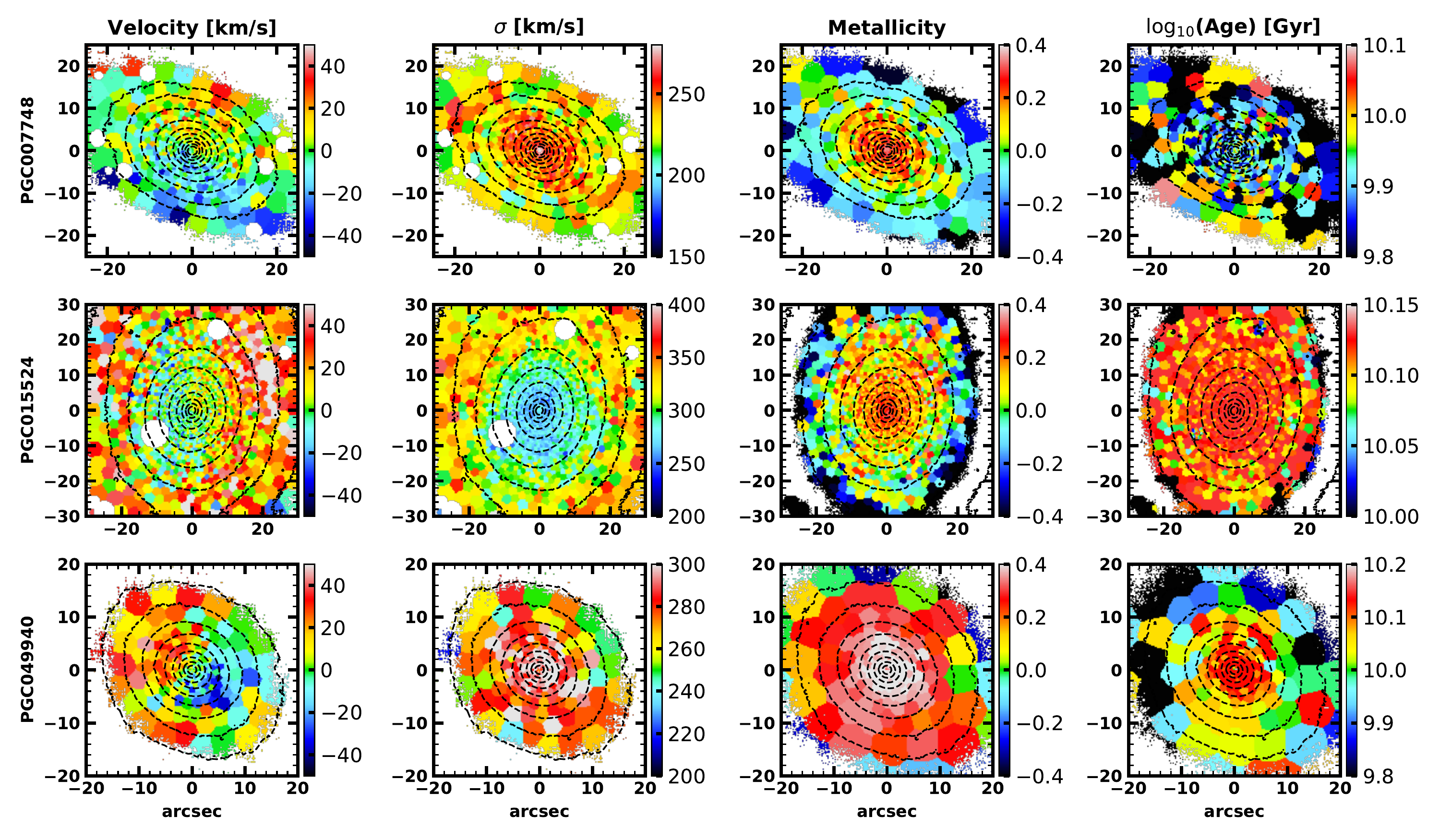}
\caption{Velocity, $\sigma$, metallicity and age MUSE maps for the studied galaxies. Black dashed contours are isophotes, plotted in steps of half a magnitude.}\label{maps}
\end{figure*}

The outlined steps produced maps of the the mean velocity and the velocity dispersion, as well as of the stellar ages and metallicities. We show these maps in Fig.~\ref{maps}. As we want to compare the velocity dispersion, ages and metallicities with the photometric radial profiles, we still needed to extract 1D information from MUSE maps. We did this using {\tt kinemetry} \citep{2006MNRAS.366..787K}, a method for analysing kinematic maps based on the ideas of iso-photometric ellipse fitting \citep[as in e.g.][]{Jedrzejewski1987}. We used the ellipse parameters (radius, flattening and the position angle) of the isophotal analysis (see Section~\ref{ss:iso}) to extract with {\tt kinemetry} the radial profiles of the velocity dispersion, age and metallicities, averaging the same areas of our galaxies as during the photometric analysis. In this way the radial profiles of photmetric and spectroscopic properties can be directly compared. In addition, we also derived the radial profiles of specific stellar angular momentum, 
\begin{equation}
    \lambda_R = \frac {\langle R |V| \rangle} {\langle R \sqrt{V^{2}+\sigma^{2}} \rangle}
\end{equation}
 \citep{Emsellem2007}, calculated as a cumulative function within an increasing aperture. We used elliptical apertures, again based on the photometric analysis (same ellipses), allowing a direct comparison with the stellar population properties, kinematics and the colour properties derived from VEGAS. Errors on the kinemetric radial profiles were obtained via error propagation. Seemingly small errors are understandable given the large number of individual bins used to estimate parameters at each radius. The formal errors on the $\lambda_R$ profiles were estimated by Monte Carlo variation of velocity and the velocity dispersion, based on the derived uncertainties on these quantities.

\section{Fit of the surface brightness profiles}\label{sec:fit}

In the recent years a considerable amount of observational works, taking advantage of the deep photometry obtained with the new generation telescopes, have demonstrated that the light profiles of many of the most massive ETGs are not well fitted by a single S{\'e}rsic law and additional components are needed \citep{Seigar2007, Donzelli2011, Arnaboldi2012, Iodice2016, Spavone2017, Spavone2018, Spavone2020, Cattapan2019, Iodice2020}.

\citet{Spavone2017} following the predictions of numerical simulations \citep{Cooper2013,Cooper2015}, described the surface brightness profiles of a sample of massive ETGs with a three component model: a S{\'e}rsic component for the innermost galaxies' regions, a second S{\'e}rsic for the central regions and an exponential (S{\'e}rsic with n=1) for the outskirts. In the simulated profiles, the first component identifies the stars formed in-situ, while the second and the third components represent the relaxed and unrelaxed accreted stars, respectively.
Since our fitting procedure is ``simulations driven'', it allows to estimate the scales at which each stellar component starts to dominate the galaxies' light profiles.

\begin{figure*}
\includegraphics[width=9cm]{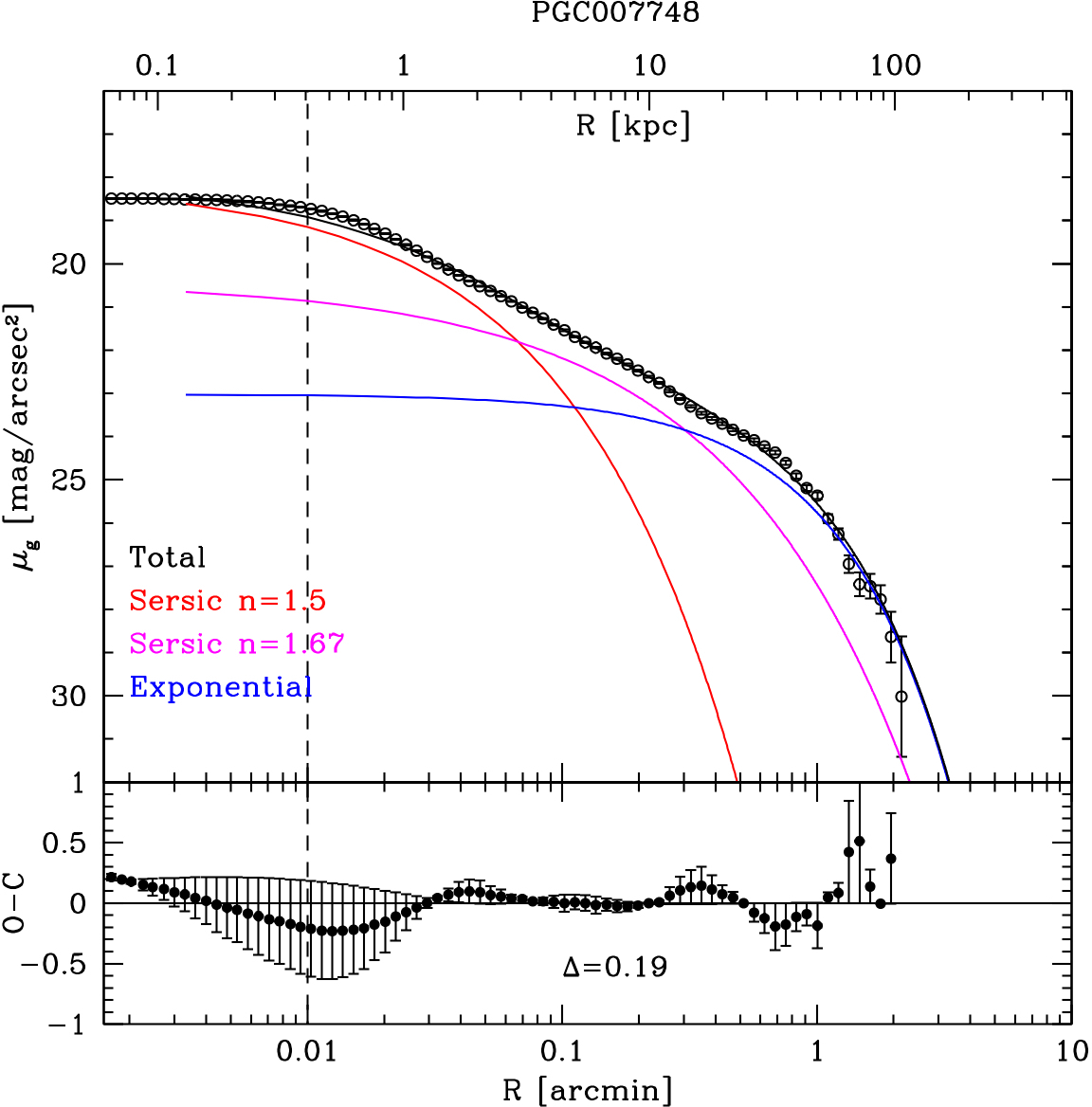}
\includegraphics[width=9cm]{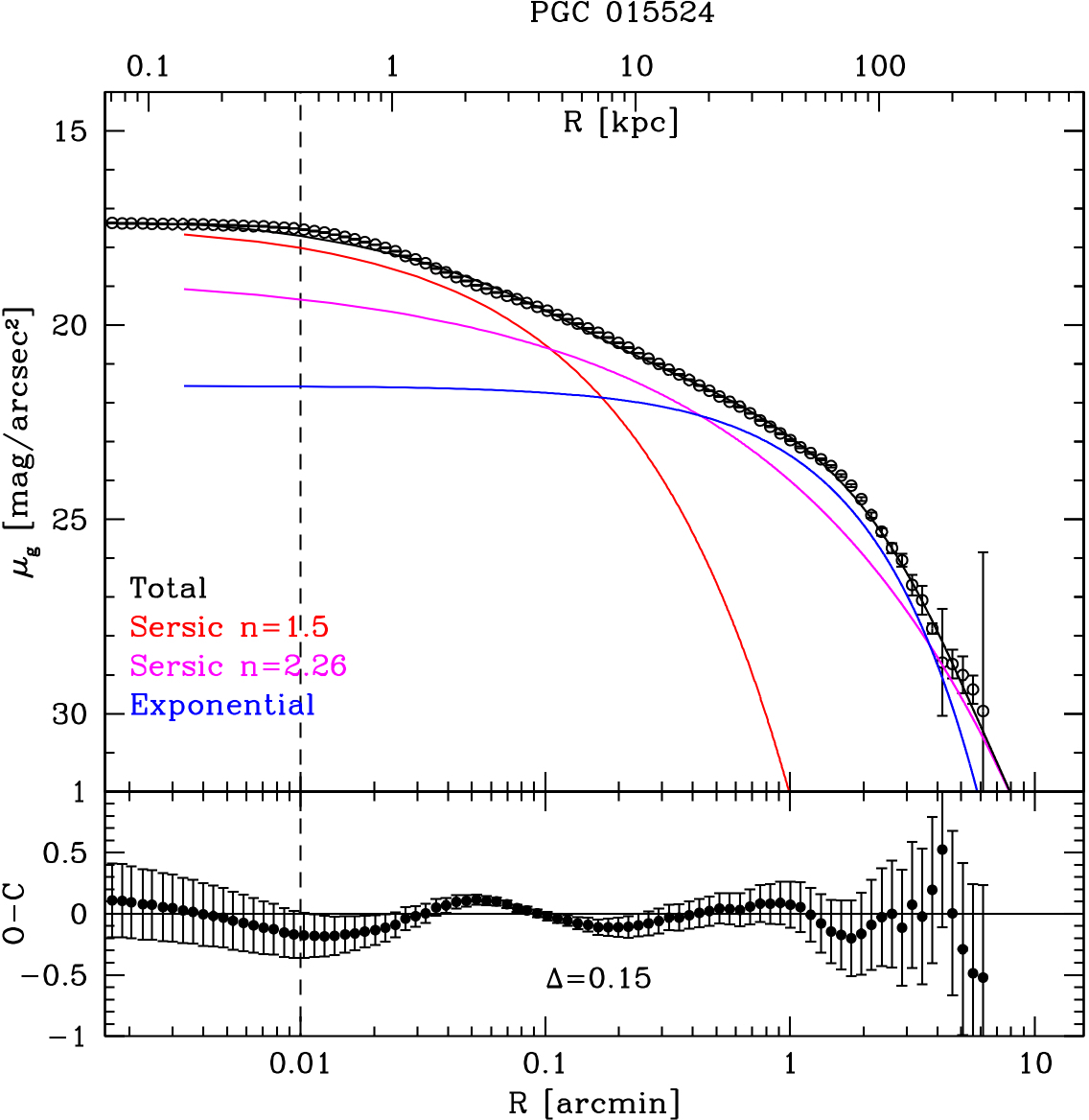}
\center \includegraphics[width=9cm]{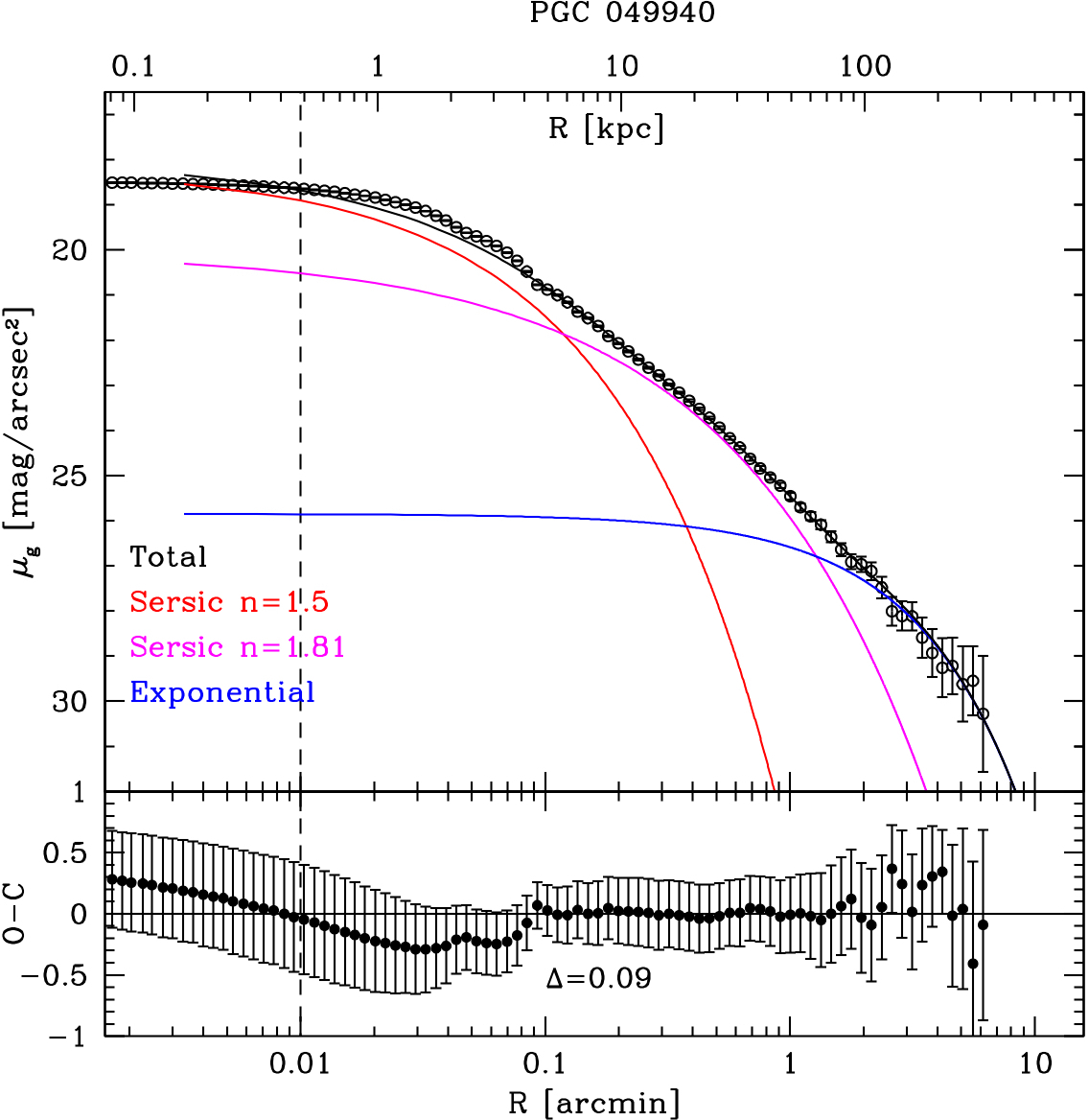}
\caption{VST {\it g} band profile of PGC007748 (top-left), PGC015524 (top-right) and PGC049940 (bottom) in linear scale, fitted with a three component
  model motivated by the predictions of theoretical simulations (see
  \citet{Spavone2017}). The bottom panels in each plot show the $\Delta$ rms scatter, obtained from the difference between the observed profiles (O) and the sum of the components from each fit (C). The dashed lines in all the panels indicate the core of the galaxies.}\label{fit}
\end{figure*}

In this work, following the procedure developed in \citet{Spavone2017}, we model the light profiles of PGC007748, PGC015524 and PGC049940 with three components fits, in order to derive hints on the balance between the different components.
Since such fits may suffer from substantial degeneracy between parameters, we adopt the typical value of the S{\'e}rsic index for the first component ($n = 2 \pm 0.5$) on the basis of the results of theoretical simulations by \citet{Cooper2013} for massive galaxies ($10^{10}\leq M_{\ast} \leq 10^{13} M_{\odot}$), to mitigate this degeneracy.
The results of the fits are shown in Fig. \ref{fit}, and the best-fitting structural parameters are reported in Table \ref{par_fit}.

\begin{table*}
\setlength{\tabcolsep}{2.1pt}
\tiny
\begin{center}
\caption{Best-fitting structural parameters for a three-component fit.} \label{par_fit}
\vspace{10pt}
\begin{tabular}{lcccccccccccccccc}
\hline\hline
Object & $\mu_{e1}$ &$r_{e1}$&$n_{1}$& $\mu_{e2}$ &$r_{e2}$&$n_{2}$&$\mu_{0}$& $r_{h}$&$R_{tr1}$&$R_{tr2}$&$R_{tr1}$&$R_{tr2}$&$f_{h,T}$\\
    &[mag/arcsec$^{2}$] &[arcesc] &  & [mag/arcsec$^{2}$] & [arcsec] & &[mag/arcsec$^{2}$] & [arcsec] &[arcsec]&[arcsec]&[kpc]&[kpc]&\\
    (1) & (2) & (3) & (4) & (5) & (6) & (7) & (8) & (9) & (10) & (11) & (12) & (13) & (14)\\
\hline \vspace{-7pt}\\
PGC007748 & 21.01$\pm$0.23  &2.79$\pm$0.18&1.5$\pm$0.9&23.69$\pm$0.07 &16.81$\pm$0.59&1.67$\pm$0.19 &23.02$\pm$0.04&23.82$\pm$0.15& 3.96&18.54&3.52&16.50&86\%$\pm$ 2\% \\
PGC015524 & 20.23$\pm$0.04  &5.22$\pm$0.10&1.5$\pm$0.5&23.19$\pm$0.07 &41.35$\pm$10.4&2.26$\pm$0.06 &21.56$\pm$0.08&36.31$\pm$7.0& 6.14&26.2&4.48&19.13&89\%$\pm$3\% \\
 PGC049940 &  21.10$\pm$0.15  &5.00$\pm$0.01&1.5$\pm$0.4&23.63$\pm$0.45 &24.36$\pm$2.50&1.81$\pm$0.05 &25.85$\pm$0.31&88.37$\pm$0.15& 7.05&75.01&5.85&62.26&77\%$\pm$5\% \\
\hline
\end{tabular}
\tablefoot{Columns 2–4 report effective magnitude, effective radius and S{\'e}rsic index for the inner component of each fit. The S{\'e}rsic index for the first component was fixed to n$\sim$2 using the models as a prior \citep{Cooper2013}. We allowed small variations of $\pm$0.5 around the mean value of n = 2. This would bracket the range of n in the simulations and allows us to obtain a better fit. Columns 5–7 list the same parameters for the second component, whereas Cols. 8 and 9 list the central surface brightness and scale length for the outer exponential component. Cols. 10-13 report the transition radii between two fit components, in arcsec and in kpc respectively, while column 14 gives the total accreted mass fraction derived from our three-component fit.}
\end{center}
\end{table*}

From this analysis we have identified two radii for each galaxy, marking the transition between the different components of the fits. These empirically defined ``transition radii'' correspond to the transition between regions dominated by different stellar populations, i.e. between in situ and accretion dominated regions of the galaxies in simulations. This transition  should be imperceptible in the azimuthally averaged profiles of ETGs (no clear inflection or break in the profiles), but it may still be detectable as a change in shape or stellar population. Since the different components are completely merged with each other at these scales, the transition from one to another is smooth and it does not happen sharply at a radius.
For this reason we have also estimated two ``transition regions'', corresponding to the range where the second and the third components of the fit starts to dominate, that is the range where the ratio between the second and the first component ($I_2/I_1$) and that between the third and the sum of the first two ($I_3/(I_1+I_2)$) pass from 50\% to 100\% (grey shaded areas in Fig. \ref{PGC7748}, \ref{PGC015524}, \ref{PGC049940} and \ref{all}).

We also derived the total mass fraction ($f_{h,T}$) enclosed in the second and the third components of our fits, and this is reported in Tab.~\ref{par_fit}. 
According to numerical simulation, this quantity has been suggested to be a proxy for the total accreted mass fraction, and it can be compared with the results of theoretical simulations as well as with other estimates for ETGs in literature (as discussed in Sec.~\ref{sec:discussion}).

\section{Results}\label{sec:SP}
\begin{figure*}
\includegraphics[width=18cm]{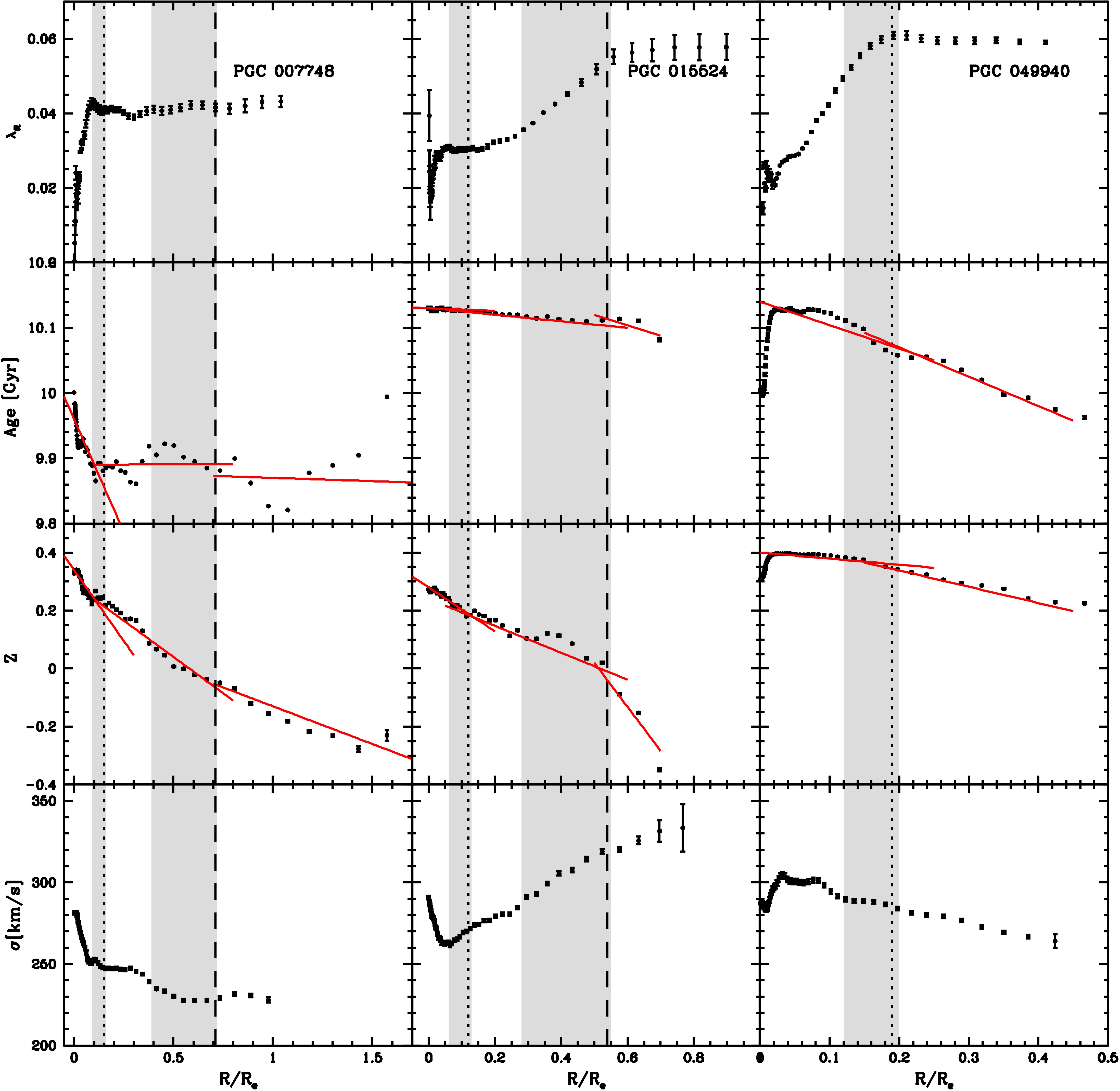}
\caption{Velocity dispersion, metallicity, age and $\lambda_R$ profiles for PGC007748 (left), PGC015524 (middle) and PGC049940 (right). The vertical dotted and dashed lines mark the position of the first and second transition radii, respectively. The grey shaded areas mark the transition regions between different components of the fit (see Sec. \ref{fit}), while the red lines are the fit performed to estimate the slopes reported in Tab. \ref{tab_slopes}}\label{all}
\end{figure*}

The main aim of this section is to compare the observables we derived from the surface photometry (i.e. light and colour distribution) with the stellar population properties derived from the spectroscopic analysis. 
In the bottom-right panels of Fig. \ref{PGC7748}, \ref{PGC015524} and \ref{PGC049940} we plot the azimuthally averaged {\it (g - r)} color profiles, while in Fig. \ref{all} we plot the velocity dispersion, the luminosity-weighted metallicity and age, and the $\lambda_R$ profile for the three galaxies studied in this paper. The transition radii derived from the fit of the surface brightness profiles as well as the transition regions, are also marked in the plots.

For PGC049940 the second transition radius occurs at $R > 60$ arcsec, that is outside the galaxy regions covered by the MUSE field of view. For PGC007748 and PGC015524 instead, both the first ($R_{tr1}$) and the second ($R_{tr2}$) transition radii are within 60 arcsec.

The azimuthally averaged {\it (g - r)} color profiles show a similar behaviour up to the first transition radius ($R_{tr2}$) for all the three galaxies (bottom panels of Fig. \ref{PGC7748}, \ref{PGC015524} and \ref{PGC049940}). 
They all have redder colors in the central regions. A gradient towards bluer colors is observed from the galaxies' centers out to the first transition radius ($R_{tr1}$). 
For $R_{tr1}\leq R\leq R_{tr2}$ the color profile of PGC007748 remains almost flat, while for PGC015524 it is slightly decreasing. The opposite happens for the color profile of PGC049940, which shows a mild increase in the region $R_{tr1}\leq R\leq R_{tr2}$. 
The three galaxies also differ in their outskirts. 
The color profile for PGC007748 is almost constant ($g-r\sim0.8$~mag) 
beyond $R_{tr2}$ up to $\sim$ 40 arcsec ($R\sim 0.7$ arcmin, $R/R_e \sim 1.5$). Even if the slope of the color profile beyond $R_{tr2}$ is consistent with flat, the data indicate a trend for a decreasing values towards bluer colours, $g-r\leq0.7$~mag at larger galactic-center distances ($R \geq 0.7$ arcmin).
The outskirts of PGC015524 show a steeper color gradients toward redder colors ($g-r\sim0.8-0.9$~mag) with respect to the average colors at smaller radii  $\leq$ 100 arcsec ($R/R_e \leq 2$). 
The color profile in the outer regions of PGC049940 ($R/R_e \geq 0.2$)
is slightly increasing toward redder colors ($g-r\sim0.85-0.95$~mag), despite the large errorbars.

It is worth noticing that, for all the three galaxies, a discontinuity and a change of slope occurs in the color profiles at the transition regions.

Inside the first transition radius, in both PGC007748 and PGC015524 the
velocity dispersion ($\sigma$) decreases (bottom row in Fig. \ref{all}). The $\sigma$ profile for PGC049940 shows a drop in the center and a plateau, and then decreases up to the first transition radius. For $R \geq R_{tr1}$, $\sigma$ is still decreasing in 
PGC007748 and PGC049940, whereas it shows a steeper positive gradient toward larger values ($\sigma \sim 280-330$~km/s) in PGC015524. In this galaxy, $\sigma$ is still growing in the outskirts at $R \geq R_{tr2}$, being $\sim350$~km/s at $R\sim1R_e$. 
In PGC007748, $\sigma$ remains almost constant ($\sim220$~km/s) at $R\geq R_{tr2}$.

The metallicity $Z$ profiles have different shape in the three galaxies (see second row from bottom in Fig.~\ref{all}).
In PGC007748, $Z$ decreases with radius, with a steeper gradient observed for $R_{tr1}\leq R \leq R_{tr2}$ and for $R\geq R_{tr2}$.
The metallicity gradient is shallower in PGC015524 than observed in 
PGC007748,  up to $R_{tr2}$, but we still observe a change in the slope 
at $R_{tr1}\leq R \leq R_{tr2}$.
In PGC049940, $Z$ remains almost constant inside $R_{tr1}$ and decreases at $R\geq R_{tr1}$, except for the large change in the innermost regions ($R/R_{e}\sim 0.02$).

Similar behaviour is found in the age profiles (see third row from bottom in Fig.~\ref{all}). In PGC007748, age decreases with radius out to $R\simeq R_{tr1}$, it varies from $\sim10$ Gyr to 9.9 Gyr, and remains almost constant outwards.
In PGC015524 the stellar population age is about 10.12 Gyr at all radii. 
In PGC049940, age is almost constant at $\sim 10.1$~Gyr for $R\leq R_{tr1}$ and decreases down to 9.9 Gyr at larger radii.

In the top panels of Fig. \ref{all} we plot the $\lambda_R$ parameter, which is a dimensionless parameter introduced by \citet{Emsellem2007} as a proxy for the baryon projected specific angular momentum. We found that all three galaxies have very low $\lambda_R$ and almost flat profiles (see eq. 12, 13 and 14 in \citealt{Schulze2020}), in agreement with their classification as slow rotators \citep{Emsellem2011}. For PGC007748 the $\lambda_R$ profile is slightly rising up to $R/R_{e} \sim 0.1$, while beyond this radius it tends to flatten. The $\lambda_R$ profiles for PGC015524 and PGC049940 are almost flat up to the second and the first transition radius respectively, where they show a peak. Beyond these radii the $\lambda_R$ profile for PGC015524 show a mild increase, while the opposite happens for PGC049940, which has a decreasing behavior.

In order to quantify the changes in the described profiles and correlate them with the transition radii from the photometry, we estimated the slopes\footnote{For PGC049940 the innermost regions ($R/R_{e}\sim 0.02$) have been excluded from the fit, since the large variations of both metallicity and age in these regions would affect the fit, leading to slopes not representative of the profiles.} of the profiles in three regions: $R \leq R_{tr1}$, $R_{tr1} < R < R_{tr2}$ and $R \geq R_{tr2}$. 
The gradients of the metallicity, $\Delta\ [Z/H]$, and age, $\Delta\ Age$, have been derived by performing fits of the metallicity and age profiles plotted against $log (R/R_{e})$. Following the definitions in \citet{Kuntschner2010}, the metallicity and age gradients are defined as:  
\begin{equation}
    \Delta\ [Z/H] = \frac{\delta\ [Z/H]} {\delta\ log R/R_{e}}
\end{equation}

\begin{equation}
    \Delta\ Age = \frac{\delta\ Age} {\delta\ log R/R_{e}}
\end{equation}

The derived values for surface brightness, metallicity and age are reported in Tab. \ref{tab_slopes}, and they will be discussed in Sec. \ref{sec:discussion}. 

The analysis presented in this section suggests that there is a quite evident correlation between the color distribution, kinematics and stellar population profiles with the transition radii setting the scale of the different components in each galaxy.

\begin{table}
\setlength{\tabcolsep}{2.5pt}
\begin{center}
\caption{Logarithmic gradients of the surface brightness, metallicity and age profiles in three regions: $R \leq R_{tr1}$, $R_{tr1} < R < R_{tr2}$ and $R \geq R_{tr2}$.} \label{tab_slopes}
\vspace{10pt}
\begin{tabular}{lcccccccccc}
\hline\hline
Parameter& $s_{R<R_{tr1}}$&$s_{R_{tr1}<R<R_{tr2}}$&$s_{R>R_{tr2}}$&\\
\hline\hline
&&PGC007748\\
\hline
$\mu_{r}$& 3.32&3.40&7.37&\\
Z&-0.07&-0.43&-0.69&\\
Age&-0.06&0.005&-0.05&\\
\hline
&&PGC015524&&&&\\
\hline
$\mu_{r}$& 2.46&3.21&6.19&\\
Z&-0.05&-0.32&-2.04&\\
Age&-2.1$\times 10^{-6}$&-0.04&-0.22&\\
\hline
&&PGC049940&&&&\\
\hline
$\mu_{r}$& 3.631&4.649&5.711&\\
Z&-0.03&-0.38&-&\\
Age&-0.03&-0.26&-&\\
\hline\hline
\end{tabular}
\end{center}
\end{table}

\section{Discussion: Observations (photometry and kinematics) versus simulations}\label{sec:discussion}

In this section we compare the results of both the photometric and kinematic analysis with the 
prediction of theoretical simulations, in order to trace the accretion history of the studied galaxies.

\subsection{Comparison with simulations and previous observational works}

\begin{figure*}
\center
\includegraphics[width=18cm]{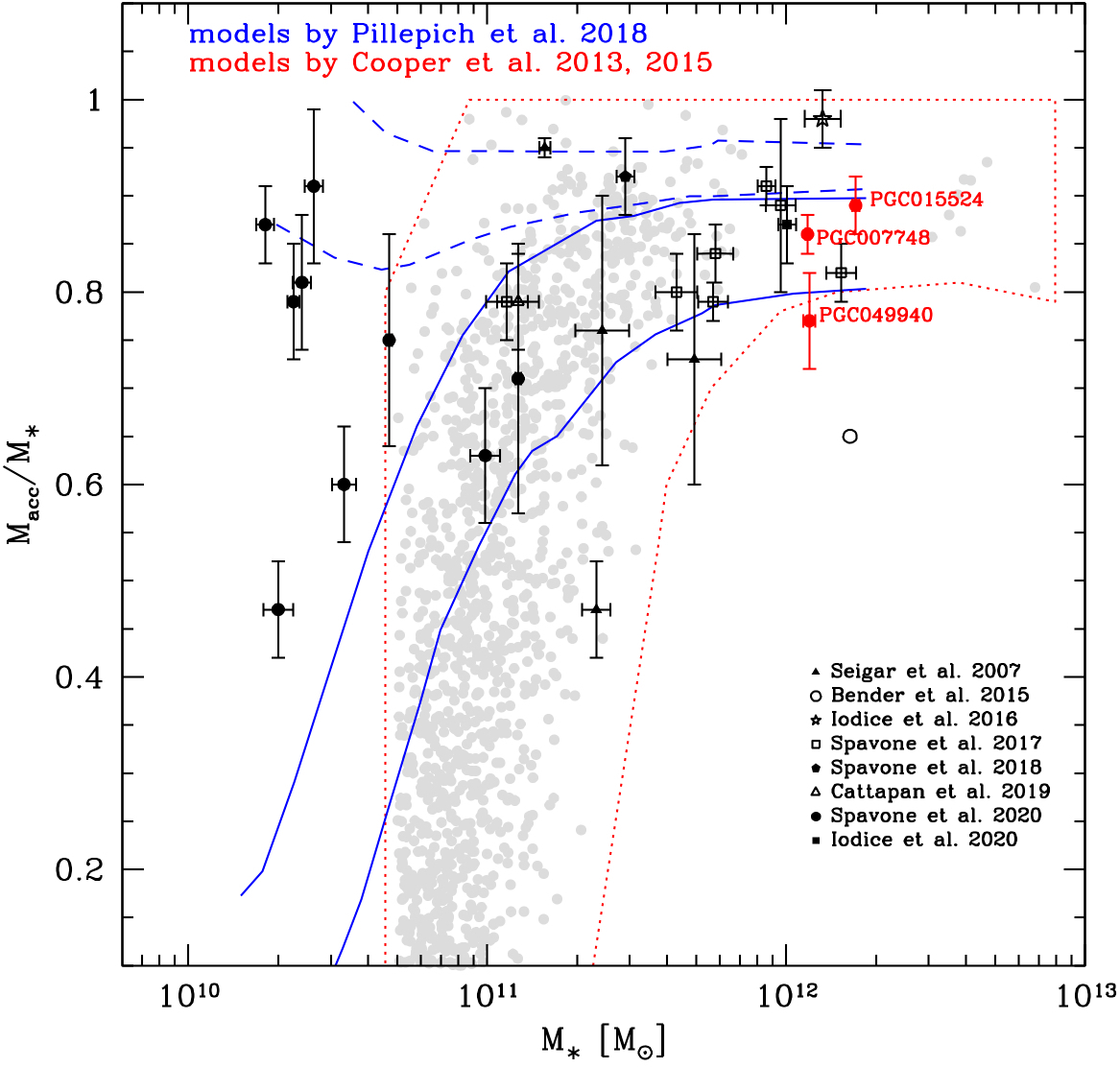}
\caption{Accreted mass fraction vs. total stellar mass for ETGs. Black symbols correspond to other BCGs from the literature
 \citep{Seigar2007, Bender2015, Iodice2016, Iodice2020, Spavone2017, Spavone2018, Spavone2020, Cattapan2019}, while red circles are for galaxies in this work. Red region
   encloses the
 predictions of cosmological galaxy formation simulations by
 \citet{Cooper2013,Cooper2015}, which are indicated as grey dots. Blue continuous and dashed regions
 indicate the accreted mass fraction measured within 30 kpc and
 outside 100 kpc, respectively, in Illustris simulations by
 \citealt{Pillepich2018} (see their Fig. 12).}\label{halo}
\end{figure*}

As described in Sec.~\ref{sec:fit}, by fitting the azimuthally averaged surface brightness distribution in the $g$ band, we can proceed and assuming the last two components represent accreted stars, estimate its mass fraction ($f_{h,T}$).
Theoretical simulations predict that this quantity is a function of the total stellar mass in the galaxy, being larger for massive galaxies \citep{Cooper2013, Cooper2015, Pillepich2018}. 
First estimates of $f_{h,T}$, based on deep imaging, are fully consistent with theoretical prescriptions \citep{Seigar2007, Bender2015, Iodice2017, Spavone2017, Spavone2018, Cattapan2019, Iodice2020, Spavone2020}.
In Fig. \ref{halo} we compare the values of $f_{h,T}$ derived for the three galaxies studied in this paper with previous estimates. In agreement with previous theoretical and observational estimates, the 
outermost components in the galaxy light distribution account for most of the total galaxy stellar mass, in the range of 80\%-90\%.

In order to test the robustness of our estimate of the accreted mass fraction, as already done by \citet{Spavone2020}, we checked how much this fraction is affected by leaving free the S{\'e}rsic index for the first component of the fit. We found that the S{\'e}rsic index ranges between 0.8 and 1, and that the accreted mass fractions change of 1\%-3\%, while the rms scatter does not change significantly.

In a recent work by \citet{Schulze2020} the authors use hydrodynamic cosmological 
{\it Magneticum Pathfinder} simulations to investigate the stellar kinematics of a sample of 
galaxies, out to 5 half-mass radius\footnote{The half-mass radius is considered to be equal to the effective radius $R_e$.}. 
\citet{Schulze2020} use the shape of the $\lambda_R$ profiles of simulated galaxies to classify them. They find three characteristic profile shapes: {\it i)} decreasing profiles, having a central peak at $R\sim0.5-2R_e$, and a decrease beyond this radius; {\it ii)} increasing profiles, which continuously increase out to the external regions; {\it iii)} flat profiles, which remains almost flat over the whole radial range.
According to \citet{Schulze2020}, galaxies with decreasing $\lambda_R$ 
profiles have a different accretion history with respect to those having 
flat or increasing profiles. The former systems acquire most of the stellar 
mass through minor mergers, differently from the galaxies with flat or 
increasing profiles, which build up from major mergers.
In addition, \citet{Schulze2020} found that the radius marking the kinematic transition provides a 
good estimate of the transition radius between the in-situ and accreted component in the 
photometric profiles. In particular, the peak of the $\lambda_R$ declining profile corresponds to 
the transition radius between the in-situ to ex-situ component.

The $\lambda_R$ profiles of galaxies in our sample are shown in the top panels of Fig.~\ref{all}. According to the classification provided by \citet{Schulze2020} the three galaxies studied in this paper have almost flat $\lambda_R$ profiles, and very low $\lambda_{max}$. All the profiles are slightly rising, and for PGC015524 and PGC049940 they show a peak corresponding to the second and first transition radius, respectively. The $\lambda_R$ profile of PGC007748 increases up to just before first transition radius, then stays roughly constant, and we don't see the decrease within the MUSE data.

Therefore, all the studied galaxies are slow rotators at the range that we probe \citep{Emsellem2007, Emsellem2011}.Moreover,
according to the predictions by \citet{Schulze2020}, the probed region (by MUSE) indicates that the mass assembly was via major mergers (also supported by the velocity maps in Fig. \ref{maps}, and see below), but we can not rule out that minor mergers are not contributing in the outskirts. In particular, PGC049940 could be a good candidate for minor mergers based on the photometry. However, the MUSE data do not go far enough out to probe the regions where the minor mergers would contribute the most, and therefore we do not know if $\lambda_R$ would change according to the prescriptions by \citet{Schulze2020}.

Additional clues on the formation pathways of investigated galaxies are found in stellar velocity presented in Fig. \ref{maps} (and \citealt{2018MNRAS.477.5327K}). PGC007784 and PGC015524 are characterised by the ``prolate-like'' rotation (rotation around the major axis), while PGC0049940 has the more common rotation around the minor axis. Assuming that ``prolate-like" rotation is a consequence of major mergers \citep[e.g.][]{2018MNRAS.473.1489L}, observed kinematics are consistent with the predictions by \citet{Schulze2020}. Another interesting difference is found between PGC015524 and PGC049940, which also have emission-line gas detected \citep{Pagotto2021}. PGC15524 has a rather disturbed and filamentary gas distribution, extending to the edge of the MUSE field, while PGC049940 has a regular nuclear disk (within 2''). These observation may point out to a somewhat different assembly history of PGC049940 from the other two galaxies.

In the theoretical work by \citet{Cook2016}, based on Illustris simulations, 
the authors perform detailed analysis of the stellar halo properties of a 
sample of ETGs by studying their surface brightness, colors and stellar 
populations. In particular, they address the accretion history of the simulated galaxies by studying the gradients of the above mentioned profiles in different galaxies' regions, 
that are $0.1R_e < R < 1R_e$ (inner galaxy), $1R_e < R < 2R_e$ (outer galaxy) and $2R_e < R < 4R_e$ (stellar halo). 
The main conclusions of their work are that, at fixed mass, the gradients of metallicity and surface brightness profiles beyond $2 R_e$ are correlated with the amount of accreted mass, while the age and color gradients are poor indicators of the accretion history. 
In particular, tracing the assembly history of galaxies with time (since z=1), \citet{Cook2016}
found that the metallicity and surface brightness profiles tend to flatten as the accretion 
of metal-rich stars increases in the galaxy outskirts.

\subsection{Interpretation of the results}
For the three galaxies studied in this paper, taking advantage of the long integration time and the large field of view of OmegaCam@VST, we have been able to map the light distribution out to the regions of the stellar halos, i.e. well beyond $2 R_e$ ($\sim 4 R_e$ for PGC007748, $\sim 6 R_e$ for PGC015524 and $\sim 10 R_e$ for PGC049940), therefore comparable with the theoretical predictions by
\citet{Cook2016}. 
The available integral-field observations with MUSE cover only a limited portion (the brightest regions) of each galaxy (out to $\sim 1R_e$), and as a consequence we cannot address any definitive
conclusion on the relation between stellar population and accretion history in the galaxy outskirts. Since we were able to derive the first transition radius from the in-situ to ex-situ component, by fitting the surface brightness profiles, we can focus on these regions (i.e. $R\geq R_{tr1}$) in the following discussion.

The slopes of the surface brightness profiles
are shallower in 
PGC015524 and PGC007748, whereas it is steeper in PGC049940 (see Fig.~\ref{fit}) and Tab.~\ref{tab_slopes}.
Therefore, according to \citet{Cook2016}, as well as to \citet{Amorisco2017}, where the slope 
of the ex-situ surface brightness profile depends on the total amount of accreted mass, we expect
that the PGC015524 and PGC007748 should have a larger amount of accreted mass in their outskirts, with respect to PGC049940.
This is indeed consistent with our independent estimate of the accreted mass fraction from the
surface brightness distribution, where PGC015524 and PGC007748 have $f_{h,T} \sim 86-89\%$, larger than the 77\% derived for PGC049940 (see Tab.~\ref{par_fit} and Fig.~\ref{halo}).

The flat $\lambda_R$ profiles for all three galaxies also indicate that the 
great majority of the mass has been acquired through major mergers \citep{Schulze2020}. In the case of PGC007748 and PGC015524 this is consistent with the high fractions of accreted mass we estimated, as well as steeper metallicity profiles beyond $R_{tr1}$. The limited extent of the spectroscopic data, however, does not constrain whether the outer halo of PGC049940 was built from the major or minor mergers. This means that the lower accreted mass fraction (77\%) for PGC049940, coupled with the steep surface brightness profile, could also indicate that its stellar halo is a result of minor mergers. In fact, simulations predict it is uncommon for a stellar halo to accrete a great amount of mass via minor mergers only \citep{Amorisco2017, Cook2016}.

\section{Summary and conclusions}\label{sec:conc}

In this paper we have presented the new deep images from the VEGAS survey of three massive 
($M_{*} \simeq 10^{12}$~M$_\odot$) galaxies from the M3G project \citep{2018MNRAS.477.5327K}:  
PGC007748, PGC015524 and PGC049940. 
The long integration time and the wide field of view of OmegaCam@VST allowed us to map the light 
and color distributions down to $\mu_g\simeq30$~mag/arcsec$^2$ and out to $\sim 2R_e$.
At this depth and distances we were able to derive the contribution of the several components
that dominate the galaxy's light, in particular the outer stellar envelope.
By fitting the surface brightness distribution for each object, we estimated the accreted 
mass fraction that contribute to the ``ex-situ'' component and we have compared this quantity with 
that predicted from simulations. 
The available integral-field observations with MUSE cover a limited portion 
 of each galaxy (the brightest regions out to $\sim 1R_e$), but, from the imaging analysis we
 found that they map the kinematics and stellar population beyond the first transition radius,
 where the contribution of the ex-situ component starts to dominate.

Main goal of this work was to correlate the scales of the different components set from the images
with the kinematics and stellar population profiles derived from the MUSE data. Results were
combined to address the assembly history of the three galaxies with the help of the theoretical predictions, and they are summarised below.

\begin{itemize}

    \item All three galaxies have a large amount of accreted mass, in the range 77\% (for
    PGC049940) to 89\% (for PGC015524). Such a large values are expected from simulations for galaxies of comparable stellar masses ($M_{*} \simeq 10^{12}$~M$_\odot$). In addition, they are also consistent with the accreted mass fraction estimated from deep imaging data available for other galaxies of similar masses (see Fig.~\ref{halo}).
    
    \item As predicted by theoretical work from \citet{Schulze2020}, there exists a correlation between the shape of the $\lambda_R$ profile with the transition radius from the in-situ to ex-situ components. In PGC007748 and PGC049940 this corresponds to the peak of the $\lambda_R$ profile (see top panels of Fig.~\ref{all}). In PGC015524 the $\lambda_R$ starts to slightly increase and it reaches its highest value at the second transition radius (where the stellar envelope dominates the accreted component).
    The observed $\lambda_R$ profiles are, however, mostly flat throughout the radial range of the MUSE data. Based on the \citet{Schulze2020} simulations, this suggests that the inner parts of galaxies were built via major mergers. Stellar velocity maps \citep{2018MNRAS.477.5327K}, where PGC007748 and PGC015524 have ``prolate-like'' rotation, also support such evolutionary scenario. In the case of PGC049940, this might not be the full picture, as its velocity map is regular and lower estimate of the accreted mass might indicate a less violent assembly history.
    
    \item The gradients of the metallicity profile inside and outside the transition radii (i.e. for the in-situ and ex-situ components) are different for the three galaxies, as also shown by the change of slope in correspondence of the transition radii. In the region between the first and the second transition radius there are no substantial differences between the metallicity gradients of the three galaxies.
    However, the ex-situ component starts to be dominant outside the second transition radius. The only galaxy for which we can discuss gradients in this region is PGC007748. The metallicity and age profiles for PGC049940 are, in contrast, not extended enough to cover the region of the second transition radius. For PGC015524 we only have few points beyond $R_{tr2}$ and the gradients in this region are not reliable enough to draw conclusions.
    In the ex-situ component of PGC007748, the metallicity profile tends to be flattened. Same behaviour is observed in the surface brightness profile, indicating that more metal-rich stars are accreted in this galaxy. This "correlation" between the metallicity gradient and the surface brightness profile with the total amount of accreted mass is consistent with simulations by \citet{Cook2016}.
    
\end{itemize}

All the above results are combined in a coherent picture tracing the assembly history of the three 
galaxies studied in this paper.
PGC049940 has a lower accreted mass fraction (77\%) then the other two galaxies. According to simulations by \citet{Schulze2020}, at the range probed by MUSE data, the flat $\lambda_R$ profile and its low amplitude, indicate that the mass assembly in this galaxy was via major mergers. However, the steep surface brightness profile could 
indicate that the stellar halo in this galaxy is assembling by minor mergers \citep{Amorisco2017, Cook2016}. For this reason, we can not rule out that minor mergers are not contributing in the outskirts of PGC049940. The higher accreted mass fraction estimated for PGC007748 and PGC015524 and the flat 
$\lambda_R$ profiles suggest that a great majority of the mass has been acquired through major 
mergers, which have also shaped the shallower metallicity profile observed at larger radii ($R\geq R_{tr1}$). Moreover, these galaxies also have prolate-like velocity maps, which is most likely a consequence of a major merger on a radial orbit.

This work represents the first observational attempt to combine imaging and spectroscopy to trace the assembly history of massive galaxies. 
Having the deep images is crucial to map the light distribution down to faintest regions 
of stellar halos and therefore to set the scales of the main galaxy components. 
The results we have obtained are encouraging and we therefore plan to acquire deep 
and more extended integral field data, which are needed to study the kinematics and stellar population content in the ex-situ component.

\begin{acknowledgements}
We are very grateful to the anonymous referee for his/her comments and suggestions which helped us to improve and clarify our work. MS and EI acknowledge financial support from the VST
project (P.I. P. Schipani).
EI acknowledges financial support from the European Union
Horizon 2020 research and innovation programme under the Marie Skodowska-Curie grant agreement n. 721463 to the SUNDIAL ITN network. DK and MdB acknowledge  financial support through the grant GZ: KR 4548/2-1 of the Deutsche Forschungsgemeinschaft. 
\end{acknowledgements}

\bibliographystyle{aa.bst}
  \bibliography{M3G}

\end{document}